\documentclass[final,5p,times,twocolumn]{elsarticle}

\usepackage{amssymb}

\usepackage{slashed}
\usepackage{bbm}
\usepackage{amsmath}
\journal{Physics Letters B}
\usepackage{hyperref}
\makeatletter
\def\ps@pprintTitle{%
 \def\@oddhead{IPPP/16/93}
 \let\@evenhead\@empty
 \let\@oddfoot\@empty%
 \let\@evenfoot\@empty}
\makeatother
\begin{document}
\begin{frontmatter}


\title{Gauge covariance\\ of the fermion Schwinger-Dyson equation in QED}




\author[wm]{Shaoyang Jia }
\ead{sjia@email.wm.edu}
\author[jlab,wm]{ M.R. Pennington }
\ead{michaelp@jlab.org}
\address[wm]{Physics Department, College of William \& Mary, Williamsburg, VA 23187, USA}
\address[jlab]{Theory Center, Thomas Jefferson National Accelerator Facility, Newport News, VA 23606, USA}

\begin{abstract}
Any practical application of the Schwinger-Dyson equations to the study of $n$-point Green's functions of a field theory requires truncations, the best known being finite order perturbation theory. Strong coupling studies require a different approach. In the case of QED, gauge covariance is a powerful constraint. By using a spectral representation for the massive fermion propagator in QED, we are able to show that the constraints imposed by the Landau-Khalatnikov-Fradkin transformations are linear operations on the spectral densities. Here we formally define these group operations and show with a couple of examples how in practice they provide a straightforward way to test the gauge covariance of any viable truncation of the Schwinger-Dyson equation for the fermion 2-point function.
\end{abstract}

\end{frontmatter}
\section{Introduction}
The natural way to study a strong coupling theory is to solve the field equations of the theory, known as the Schwinger-Dyson equations. Solving such equations provides a nonperturbative approach to QCD with applications to hadronic physics \cite{Bashir:2012fs}. Since this is an infinite system of coupled integral equations, their solution for any particular Green's function, such as the fermion propagator we consider here, requires a truncation of this infinite system. In practice, when studying the fermion propagator  this means making an ansatz for the fermion-boson vertex. As a guide for QCD, here we deduce the constraints required on such structures that gauge covariance in QED imposes. Considering arbitrary dimensions allows us to make connections between three and four dimensional theories, which are of current interest. The fermion propagator in QED is expected to have a simple analytic structure with poles that correspond to physical particles, like the electron, and with branch cuts corresponding to particle creation such as additional photons, or electron-positron pairs. Such analytic structure motivates a spectral representation for the fermion propagator \cite{peskin1995introduction}. This turns out to be particularly useful for realizing the constraints of gauge covariance. Here we restrict attention to covariant gauges for ease of calculation.

The relation between QED Green's functions evaluated in different covariant gauges is specified by the Landau-Khalatnikov-Fradkin transformation (LKFT) \cite{Zumino:1959wt}. Differential forms of LKFT are called Nielson identities. A review of the LKF transform and Nielson identities can be found in the introductory paragraphs in Section 11 of Ref.~\cite{Bashir:2004mu}. Derivations of LKFT using BRST transforms or employing functional derivative with up-to-date conventions, can be found in Ref.~\cite{Sonoda:2000kn}, while for Nielson identities, see Refs.~\cite{Breckenridge:1994gs,PhysRevD.62.076002}. In this article we explore the general gauge covariance requirement imposed on the Schwinger-Dyson equation (SDE) for massive fermions that is independent of the solution in one particular gauge.

This article is organized as the following. In Section \ref{ss:LKFT}, the spectral representation for the fermion propagator is introduced to deduce the exact solutions to the LKFT for the fermion propagator. In Section \ref{ss:consitency}, the consistency requirement between SDE and the LKFT for the fermion propagator has been proposed. Meanwhile, two examples are included to explain how identities previously formulated in this article work in practice. Section \ref{ss:conclusion} gives the conclusion.
\section{LKFT for fermion propagator in spectral representation\label{ss:LKFT}}
\subsection{Spectral representation of fermion propagator\label{(ss:spectral_rep)}}
\begin{figure}
\centering
\includegraphics[width=0.75\linewidth]{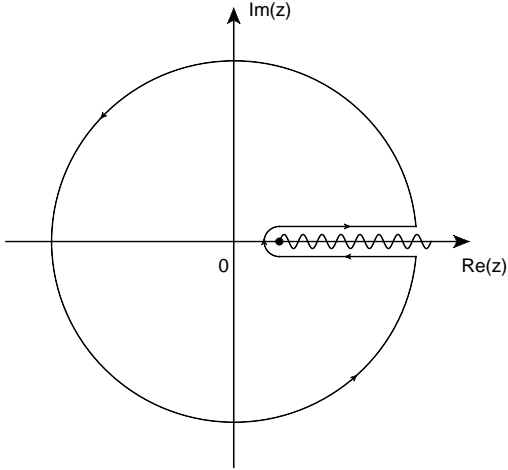}
\caption{The illustration of analytic functions with branch cuts along the positive real axis, corresponding to the production of real particles that can only be achieved in the timelike region. We use the Bjorken-Drell metric therefore this happens when $z=p^2/s>0$. The contour can be used to prove Eq.~\eqref{eq:fermion_spectral_rep} using the Cauchy integral formula with $z$ replaced by $p^2$.}
\label{fig:analytic_fz}
\end{figure}

The existence of spectral representations for fermion propagators relies on the exact analytic structure of propagator functions in the complex momentum plane. For massive fermions in QED, we assume singularities of their propagator functions can only consist of branch cuts along the positive real axis with, in addition, a finite number of poles, while being  holomorphic everywhere else in the complex momentum plane. Fig.~\ref{fig:analytic_fz} sketches this type of function with only branch-cut singularities illustrated.\footnote{Poles correspond to summing up free-particle propagators with different mass, the value of which could be complex. Since they are trivial to include, they are not shown in Fig. \ref{fig:analytic_fz}.}
The fermion propagator carrying momentum $p$, $S_F(p)$, has two Dirac structures identified as the coefficients of the $\gamma$-matrices and the identity matrix: $S_F(p)=S_1(p^2)\slashed{p}+S_2(p^2)\mathbbm{1}$.
We can then associate a spectral function $\rho_{j}$ to each of these scalar functions;
\begin{equation}
\rho_{j}(s;\xi)\;=\;-\dfrac{1}{\pi}\;\mathrm{Im}\big\{S_{j}(s+i\varepsilon;\xi)\big\}\, ,\label{eq:fermion_inv_spectral_rep}
\end{equation}
so that when $j=1,2$  the $\rho_{j}$ are the discontinuities across the branch cut in Fig.~1 for $S_1$ and $S_2$ respectively. Pole terms are implicitly included by using the Feynman $i\varepsilon$ prescription. Then provided the propagator functions go to zero as $|p^2|\rightarrow \infty$, they are completely given by 
\begin{equation}
S_{j}(p^2;\xi)\;=\;\int_{m^2}^{+\infty}ds\;\dfrac{\rho_{j}(s;\xi)}{p^2-s+i\varepsilon}\, ,\label{eq:fermion_spectral_rep}
\end{equation}
using standard Cauchy integration. Subtractions may be required to ensure convergence.\footnote{That the integrals defined by Eq.~\eqref{eq:fermion_spectral_rep} converge without the need for subtractions is assured
by the renormalizability of QED in $d < 4$ dimensions.}
Note the dependence of the fermion propagator on the covariant gauge fixing parameter $\xi$ has been made explicit, as this is crucial for the LKF transformation. 
One would expect in a massive fermion theory that the spectral functions have components that are delta functions, corresponding to particles of definite mass, and a series of theta functions at each particle production threshold. However, other structures may be required from the solution of the fermion SDE and its gauge covariance, as we will comment on below.

\subsection{LKFT as group transformations}
LKFT specifies how Green's functions change from one gauge to another. For the fermion propagator, LKFT was originally formulated in coordinate space: \cite{Zumino:1959wt}
\begin{equation}
S_F(x-y;\xi)=\exp\left\{ie^2\xi\left[M(x-y)-M(0)\right]\right\}\,S_F(x-y;0),\label{eq:LKFT_SF}
\end{equation}
where $S_F(x-y;\xi)$ is the propagator calculated in any covariant gauge.
$M(z)=-\int d\underline{l}~e^{-il\cdot z}/(l^4+i\varepsilon)$, where the integral measure is $d\underline{l}=d^d l/(2\pi)^d$.
Differentiating Eq.~\eqref{eq:LKFT_SF} with respect to $\xi$ and then taking the Fourier transform, one obtains
\begin{equation}
\dfrac{\partial}{\partial\xi}S_F(p)=ie^2\int d\underline{l}\,\dfrac{1}{l^4+i\varepsilon}\,\left[S_F(p)-S_F(p-l)\right].\label{eq:Dxi_SF}
\end{equation}
The absence of dimension-odd operators in Eq.~\eqref{eq:LKFT_SF} decouples the Dirac scalar and vector components, in contrast to the SDE we consider later. Because Eq.~\eqref{eq:LKFT_SF} has no knowledge of the propagator in the initial gauge, the differential form of LKFT written in Eq.~\eqref{eq:Dxi_SF} is equivalent to its finite form. Therefore the LKFT for fermion propagator in momentum space becomes effectively a one-loop integral. 

To understand the mathematical properties of LKFT, we start with the observation that Fourier transforms are bijective. 
This implies the relation between propagators in coordinate space and propagator spectral functions is bijective as well, as established in Ref.~\cite{LKFT.fermion.epsilon}. 

Based on Eq.~\eqref{eq:LKFT_SF}, the LKFT for the fermion propagator in coordinate space is simply a real phase factor. Moreover, when considered as a linear transformation of functions in  coordinate space, the LKFT can be viewed as a group transformation. One can easily verify that when the group multiplication is defined as a function multiplication, all the requirements of group transformations are satisfied. Meanwhile, LKFT for momentum space propagators as well as for propagator spectral functions should all be group transforms, based on {\it one-to-one} and {\it onto} correspondences. 
In fact, the coordinate space representation, the momentum space representation and the spectral representation of LKFT are isomorphic representations of the same group. Additionally, since $\xi$ parameterizes the LKFT as a continuous group, the starting gauge of LKFT does not matter; only the difference in $\xi$ enters into calculation. The default initial gauge for LKFT is conveniently chosen to be the Landau gauge. For calculation with initial gauge parameter $\xi_0$, one can replace Landau gauge quantities by those at $\xi_0$ and replace $\xi$ by $\xi-\xi_0$.

For the particularly interesting spectral representation of the LKFT for fermion propagator, we have established that the LKFT is a group transformation. However, instead of simply being a phase factor, we expect the LKFT for the spectral representation to be more complicated, but still consists of linear operations. Consequently, without loss of generality 
we can write 
\begin{equation}
\rho_{j}(s;\xi)\,=\,\int ds' \,{\cal K}_{j}(s,s';\xi) \,\rho_{j}(s';0)\, ,\label{eq:LKFT_linearity_spectral_rep}
\end{equation}
where distributions ${\cal K}_{j}(s,s';\xi)$ being the LKFT in spectral representation, specify linear operations that encode $\xi$ dependences of $\rho_{j}(s;\xi)$.

In order to find out these ${\cal K}_{j}$, we could first substitute Eq.~\eqref{eq:fermion_spectral_rep} into Eq.~\eqref{eq:Dxi_SF} and complete the loop integral. Because $\rho_{j}(s;0)$ is arbitrary as far as LKFT is concerned, subsequently substituting Eq.~\eqref{eq:LKFT_linearity_spectral_rep} in gives
\begin{equation}
\dfrac{\partial}{\partial\xi}\int ds\,\dfrac{{\cal K}_{j}(s,s';\xi)}{p^2-s+i\epsilon}\,=\,-\dfrac{\alpha}{4\pi}\int ds\,\dfrac{\Xi_{j}(p^2,s)}{p^2-s+i\epsilon}\,{\cal K}_{j}(s,s';\xi)\, ,\label{eq:LKFT_k12}
\end{equation}
where the $\Xi_{j}(p^2,s)$ are determined by the effective one-loop integral, which can be evaluated using Feynman parameterization and dimensional regularization. Integrals in $d$-dimensions are traditionally performed by first making a Wick rotation and using the resulting $\,d$-fold spherical symmetry to perform the angular integrals before the radial integral \cite{peskin1995introduction}. However, one can instead perform the integration wholly in Minkowski space, by first integrating to infinity over the time component of the loop momentum and continuing the number of space dimensions to $d-1$. Here as all aspects of the integrals are known, one can readily see the results with or without Wick rotation are identical. This is a virtue of assuming a  spectral representation when all loop integrals involve only explicitly known functions. 

Eq.~\eqref{eq:LKFT_k12} is most easily solved by substituting in the following test solutions
\begin{equation}
{\cal K}_{j}=\exp\left(-\dfrac{\alpha\xi}{4\pi}\Phi_{j}\right)\, ,\label{eq:k_exponential}
\end{equation}
where distributions $\Phi_{j}$ are independent of $\xi$. The exponential of a distribution is given by definition
\begin{equation}
\exp\big\{\lambda\Phi\big\}=\sum_{n=0}^{+\infty}\dfrac{\lambda^n}{n!}\Phi^n\;=\;\delta(s-s')\,+\,\lambda\Phi+\dfrac{\lambda^2}{2!}\Phi^2+\dots\, ,
\end{equation}
with distribution exponentiation defined as $\Phi^0(s,s')\,=\,\delta(s-s')$\ and \[\Phi^n(s,s')\,=\,\int ds''\Phi(s,s'')\,\Phi^{n-1}(s'',s')\quad (n\geq 1).\] One can easily verify that ${\cal K}_{j}$ given by Eq.~\eqref{eq:k_exponential} indeed satisfy Eq.~\eqref{eq:LKFT_k12} with initial conditions ${\cal K}_{j}(s,s';0)\,=\,\delta(s-s')$ provided the distributions $\Phi_{j}$ solve the following identities\footnote{One could also use group properties to deduce Eq.~\eqref{eq:k_exponential} from the differential equations themselves. See Ref.~\cite{LKFT.fermion.epsilon} for details.}
\begin{equation}
\int ds\,\dfrac{\Phi_{j}(s,s')}{p^2-s+i\epsilon}\,=\,\dfrac{\Xi_{j}(p^2,s')}{p^2-s'+i\epsilon}\,.\label{eq:Phi_Xi}
\end{equation}
Here by writing down Eq.~\eqref{eq:Phi_Xi} the idea of a spectral representation for propagators has been generalized to express identities for other distributions. To solve Eq.~\eqref{eq:Phi_Xi}, we need to find out the linear transform acting only on the spectral variable $s$ of the free-particle propagator $(p^2-s+i\varepsilon)^{-1}$ to create any $p^2$ dependences in $\Xi_{j}(p^2,s)/(p^2-s+i\varepsilon)$.

Up until now we've applied the group nature of LKFT to reduce the $\xi$ dependence of fermion propagator spectral functions to Eq.~\eqref{eq:Phi_Xi}.This is the equation that the distributions $\Phi_{j}$ have to satisfy. To solve for $\Phi_j$, new tools will be developed in the following subsection.

\subsection{Dimensional regularization of LKFT and solutions with fractional calculus}
Utilizing well established perturbative techniques, we can calculate the functions $\Xi_{j}(p^2,s)$ from Eq.~\eqref{eq:Dxi_SF}. 
Explicitly,
\begin{align}
\dfrac{\Xi_1}{p^2-s}& = \dfrac{\Gamma(\epsilon)}{s}\left(\dfrac{4\pi\mu^2}{s}\right)^\epsilon\dfrac{-2}{(1-\epsilon)(2-\epsilon)}~_2F_1(\epsilon+1,3;3-\epsilon;z)\label{eq:Xi_1_reduced}\\[3mm]
\dfrac{\Xi_2}{p^2-s}& = \dfrac{\Gamma(\epsilon)}{s}\left(\dfrac{4\pi\mu^2}{s}\right)^\epsilon\dfrac{-1}{1-\epsilon}~_2F_1(\epsilon+1,2;2-\epsilon;z)\, ,\label{eq:Xi_2_reduced}
\end{align}
where $z=p^2/s$ and the number of spacetime dimension\footnote{We use $\varepsilon$ to denote the Feynman prescription of momentum space propagators and $\epsilon$ as how close the number of spacetime dimensions is to $4$.} is given by $d=4-2\epsilon$. One can verify by applying Eq.~(15.3.6) of Ref. \cite{abramowitz1964handbook} that hypergeometric functions in Eqs.~(\ref{eq:Xi_1_reduced},~\ref{eq:Xi_2_reduced}) are more singular than the free-particle propagator when $\epsilon>0$ in the $z\rightarrow 1$ limit. The best way to regularize them is to keep the number of spacetime dimensions explicit throughout the entire calculation.

To generate these hypergeometric functions from the free-particle propagator as implied by Eq.~\eqref{eq:Phi_Xi} using only linear operations on the spectral variable $s$ for any $\epsilon$, \lq\lq exotic'' linear operators are expected. The first clue in finding $\Phi_{j}$ from Eq.~\eqref{eq:Phi_Xi} with $\Xi_{j}$ given by Eqs.~(\ref{eq:Xi_1_reduced},~\ref{eq:Xi_2_reduced}) is realizing that the Taylor expansion in $z=p^2/s$ of the free-particle propagator is simply a geometric series. Notice that $~_2F_1(1,b;b;z)=(1-z)^{-1}$, while hypergeometric series are natural generalizations of geometric series. For integer orders of derivative, to generate any hypergeometric $~_2F_1$ linearly from the free-particle propagator, we could directly apply Eqs.~(15.2.3,~15.2.4) from Abramowitz and Stegun \cite{abramowitz1964handbook}. One natural way to generalize these differentiation formulae to accommodate fractional parameters is to use the following definition of Riemann-Liouville fractional calculus \cite{Riemann:fractional} :
\begin{equation}
I^\alpha f(z)=\dfrac{1}{\Gamma(\alpha)}\int_{0}^{z}dz'(z-z')^{\alpha-1}f(z').\label{eq:def_Riemann_Liouville_I}
\end{equation}
For $\alpha>0$, the Riemann-Liouville fractional derivative is defined as
\begin{equation}
D^\alpha f(z)=\left(\dfrac{d}{dz}\right)^{\lceil \alpha \rceil}I^{\lceil\alpha\rceil-\alpha}f(z),
\end{equation}
where $\lceil \alpha \rceil$ is the ceiling function. It follows that ${D^\alpha z^\beta =(1-\alpha+\beta)_\alpha z^{-\alpha+\beta}}$, with Pochhammer symbol defined as ${(1-\alpha+\beta)_\alpha=\Gamma(1+\beta)/\Gamma(1-\alpha+\beta)}$. With these definitions of calculus operators at fractional orders, one can then easily verify 
\begin{align}
& D^\alpha z^{a+\alpha-1}~_2F_1(a,b;c;z)\,=\,(a)_\alpha z^{a-1}~_2F_1(a+\alpha,b;c;z)\, ,\label{eq:Da2F1}\\[3mm]
& D^\alpha z^{c-1}~_2F_1(a,b;c;z)\,=\,(c-\alpha)_\alpha z^{c-\alpha-1}~_2F_1(a,b;c-\alpha;z)\, ,\label{eq:Dc2Fa}
\end{align}
as the desired generalization of Eqs.~(15.2.3,~15.2.4) of Ref.~\cite{abramowitz1964handbook}. Equipped with Eqs.~(\ref{eq:Da2F1},~\ref{eq:Dc2Fa}), Eq.~\eqref{eq:Phi_Xi} can be solved by
\begin{equation}
\phi_n\,=\,\Gamma(\epsilon)\left(\dfrac{4\pi\mu^2}{p^2}\right)^\epsilon \dfrac{\Gamma(1-\epsilon)}{\Gamma(1+\epsilon)}\,z^{2\epsilon+2-n}D^\epsilon z^{n-1}D^\epsilon z^{\epsilon-1}.\label{eq:def_phi}
\end{equation} 
where operators $\phi_n$ are defined such that at the operator level $\int ds'\Phi=\phi$. When acting on the free-particle propagator,
\begin{align}
-\phi_n z~_2F_1(1,n;n;z)=&\Gamma(\epsilon)\,\left(\dfrac{4\pi\mu^2}{p^2}\right)^\epsilon\, \dfrac{-\Gamma(2-\epsilon)}{(1-\epsilon)\Gamma(1+\epsilon)}\;\times\nonumber\\
& \hspace{1.8cm} z^{2\epsilon+2-n}\,D^\epsilon\, z^{n-1}\,D^\epsilon \,z^{\epsilon},
\end{align}
which produces the linear transforms required to generate $\Xi_{j}(p^2,s)/(p^2-s+i\varepsilon)$ from the free-particle propagator with $n=3$ for $j=1$ in Eq.~\eqref{eq:Xi_1_reduced} and $n=2$ for $j=2$ in Eq.~\eqref{eq:Xi_2_reduced}. 

Until now we have formally solved the $\xi$ dependence of fermion propagator spectral functions with arbitrary dimensions as long as hypergeometric functions are well defined. However the exponential of distributions given by Eq.~\eqref{eq:k_exponential} remains illusive. While the action of a linear operator on the propagator is completely specified once we know how it works on $z^\beta$ with arbitrary real $\beta$.
For $\phi_n$ defined by Eq.~\eqref{eq:def_phi}, we find that
\begin{equation}
{\cal K}_j\,z^\beta\, =\,\sum_{m=0}^{+\infty}\,\dfrac{(-\overline{\alpha})^m}{m!}\,\dfrac{\Gamma(n+\beta+(m-1)\epsilon-1) \Gamma(\beta+m\epsilon)}{\Gamma(n+\beta-\epsilon-1)\Gamma(\beta)}\,z^{\beta+m\epsilon},\label{eq:kn_zbeta}
\end{equation}
where $\overline{\alpha}=(\alpha\xi/4\pi)\left(4\pi\mu^2/p^2\right)^\epsilon\Gamma(\epsilon)\Gamma(1-\epsilon)/\Gamma(1+\epsilon)$ and again with $n=3,2$ for $j=1,2$. With Eq.~\eqref{eq:kn_zbeta}, actions of LKFT on spectral variables are explicit.

\section{Consistency requirement from LKFT on the fermion propagator SDE\label{ss:consitency}}
\subsection{SDE for fermion propagator spectral functions}
\begin{figure}
\centering
\includegraphics[width=0.9\linewidth]{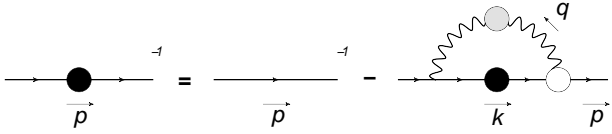}
\caption{The SDE for fermion propagator functions. Black circles correspond to connected diagrams, while the white circle stands for the one-particle irreducible (1PI) vertex. In the quenched approximation, the gray circle gets removed, while in the unquenched case it represents the connected diagram.}
\label{fig:DSE_fermion_ori_photon_mod}
\end{figure}

\begin{figure}
\centering
\includegraphics[width=0.9\linewidth]{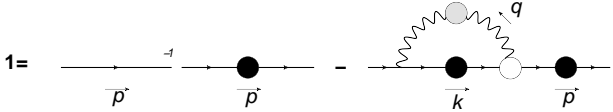}
\caption{The SDE fermion propagator linear in spectral functions obtained by multiplying $S_F(p)$ to the right of Fig.~\ref{fig:DSE_fermion_ori_photon_mod}. Black circles correspond to connected diagrams, while the white circle stands for the one-particle irreducible (1PI) vertex. In the quenched approximation, the gray circle gets removed, while in the unquenched case it represents the connected diagram.}
\label{fig:DSE_fermion_rho_photon_mod}
\end{figure}

The SDE for the fermion propagator in momentum space is represented by the diagrammatic identity in Fig.~\ref{fig:DSE_fermion_ori_photon_mod}. The SDE written in this form is most convenient for solving propagator functions directly in the spacelike region. However each diagram is not linear in spectral functions $\rho_{j}(s;\xi)$. Alternatively, multiplying $S_F(p)$ to the right gives the equivalent identity shown in Fig. \ref{fig:DSE_fermion_rho_photon_mod}. The first diagram on the right-hand side is clearly linear in $\rho_{j}(s;\xi)$. The dependence of the last diagram on the right-hand side on $\rho_{j}$ can be judged from the well-known Ward identity of QED \cite{PhysRev.78.182} .
Because the fermion-photon vertex structure $S_F(k)\,\Gamma^\mu(k,p)\,S_F(p)$ is required to share its renormalization constant with $S_F(p)$ to ensure $Z_1=Z_2$. Consequently both of them must be linear in $\rho_{j}(s;\xi)$.

Under this linear assumption, let us imagine the dependence of $S_F(k)\,\Gamma_\mu(k,p)\,S_F(p)$ on the fermion propagator spectral function $\rho_{j}(s)$ is known. After evaluating the loop integral in Fig.~\ref{fig:DSE_fermion_rho_photon_mod}, the remaining operations on spectral functions can only be linear. Taking the imaginary part of the identity in Fig.~\ref{fig:DSE_fermion_rho_photon_mod} means taking the discontinuity across the cut in Fig.~1. In the case of quenched QED the only contributions from particle production come from the fermion plus photons. We then have the coupled equations:
\begin{equation}
\begin{pmatrix}
\rho_1(s;\xi) \\ 
\rho_2(s;\xi)
\end{pmatrix}+
\int ds'
\begin{pmatrix}
\Omega_{11}(s,s';\xi) & \Omega_{12}(s,s';\xi) \\ 
\Omega_{21}(s,s';\xi) & \Omega_{22}(s,s';\xi)
\end{pmatrix} 
\begin{pmatrix}
\rho_1(s';\xi) \\ 
\rho_2(s';\xi)
\end{pmatrix}=
\begin{pmatrix}
0 \\ 
0
\end{pmatrix} .\label{eq:SDE_fermion_rho_itg}
\end{equation}
While the structure of Eq.~\eqref{eq:SDE_fermion_rho_itg} is general, the actual form of the $\Omega_{ij}$ implicitly depends on the photon propagator. In the quenched case, there is no other dependence on $\rho_j$. However, when the photon is unquenched, the $\Omega_{ij}$ contain implicit $\rho_j$ dependence through the vacuum polarization. The fact that this polarization is gauge independent is ensured by the particular gauge dependence of $\rho_j$ in Eqs.~(\ref{eq:LKFT_linearity_spectral_rep},~\ref{eq:k_exponential}) as discussed in more detail in \cite{GC.propagators.SDEs}. 

Eq.~\eqref{eq:SDE_fermion_rho_itg} is homogeneous in $\rho$ because the real inhomogeneous constant on the left-hand of the identity in Fig.~\ref{fig:DSE_fermion_rho_photon_mod} vanishes after taking the imaginary part. We will derive the explicit structure of the $\Omega_{ij}$ in the case of quenched QED below. While with unquenched photons, the $\Omega_{ij}$ are expected to include additional $\theta$-functions corresponding to other real production thresholds. Nevertheless, the general form of SDE for fermion propagator spectral functions still takes that of Eq.~\eqref{eq:SDE_fermion_rho_itg}. 
Analytic structures of the fermion-photon vertex are also subsumed into the formalism of Eq.~\eqref{eq:SDE_fermion_rho_itg} because the Ward-Green-Takahashi identity ensures the discontinuity of $S_F\Gamma^\mu S_F$ be linear in $\rho$. In general with any number of spacetime dimensions, the $\Omega(s,s';\xi)$ are distributions rather than simple functions of spectral variables $s$ and $s'$. 

In analogy with matrices being finite dimensional linear operators on vectors, distributions can be viewed as linear operations on functions that are the infinite dimensional version of vectors. Therefore for notational brevity, Eq.~\eqref{eq:SDE_fermion_rho_itg} can be written as
\begin{equation}
\begin{pmatrix}
\rho_1^\xi \\[0.5em]
\rho_2^\xi
\end{pmatrix} +
\begin{pmatrix}
\Omega_{11}^\xi & \Omega_{12}^\xi \\[0.5em] 
\Omega_{21}^\xi & \Omega_{22}^\xi
\end{pmatrix} 
\begin{pmatrix}
\rho_1^\xi \\[0.5em] 
\rho_2^\xi
\end{pmatrix}=
\begin{pmatrix}
0 \\[0.5em]
0
\end{pmatrix},\label{eq:fermion_prop_SDE_abs}
\end{equation}
where distributional multiplications are understood with the integrals over spectral variables being implicit, adopting a similar convention to matrix multiplication.

\subsection{The general result}
The linear operator $\Omega$ in Eq.~\eqref{eq:fermion_prop_SDE_abs} is determined by the interactions of QED, specifically the fermion-photon vertex. Without knowing the vertex exactly, one needs to come up with an ansatz to truncate the infinite tower of SDEs. Such an ansatz determines $\Omega$, which after solving Eq.~\eqref{eq:fermion_prop_SDE_abs}, subsequently determines the spectral functions $\rho_{j}(s;\xi)$. With an arbitrary ansatz, the $\rho_{j}(s;\xi)$ solved from Eq.~\eqref{eq:fermion_prop_SDE_abs} in different gauges are not necessarily related by the LKFT. Since the $\xi$ dependence of $\rho_{j}$ is known exactly, a natural question arises is what is the requirement on $\Omega$ such that solutions to Eq.~\eqref{eq:fermion_prop_SDE_abs} satisfies LKFT. 

To answer this question, let us start by substituting ${\rho_{j}^\xi={\cal K}_{j}^\xi\rho_{j}^0}$, the abstract version of Eq.~\eqref{eq:LKFT_linearity_spectral_rep}, into Eq.~\eqref{eq:fermion_prop_SDE_abs}.
Noting that $(\mathrm{diag}\{{\cal K}_1^\xi,~{\cal K}_2^\xi \})^{-1}=\mathrm{diag}\{{\cal K}_1^{-\xi},~{\cal K}_2^{-\xi} \}$ defines this distribution inversion, we arrive at our final result 
\begin{equation}
\begin{pmatrix}
\Omega_{11}^0 & \Omega_{12}^0 \\[0.5em] 
\Omega_{21}^0 & \Omega_{22}^0
\end{pmatrix} =
\begin{pmatrix}
{\cal K}_1^{-\xi} & \\[0.5em]
 & {\cal K}_2^{-\xi}
\end{pmatrix} 
\begin{pmatrix}
\Omega_{11}^\xi & \Omega_{12}^\xi \\[0.5em] 
\Omega_{21}^\xi & \Omega_{22}^\xi
\end{pmatrix} 
\begin{pmatrix}
{\cal K}_1^\xi & \\[0.5em]
 & {\cal K}_2^\xi
\end{pmatrix},\label{eq:consistency_fermion_prop_SDE_LKFT}
\end{equation}
or more compactly as $\Omega_0={\cal K}_{-\xi}\Omega_\xi {\cal K}_{\xi}$. One can also prove that $\Omega$ satisfying Eq.~\eqref{eq:consistency_fermion_prop_SDE_LKFT} will produce $\rho_{j}$ with the correct $\xi$ dependence given by LKFT \cite{LKFT.fermion.epsilon}. Therefore Eq.~\eqref{eq:consistency_fermion_prop_SDE_LKFT} is the necessary and sufficient condition for the LKFT and the SDE for the fermion propagator to be consistent with each other. 
\subsection{Two simple applications of the general result}
We consider here two examples of applying Eq.~\eqref{eq:consistency_fermion_prop_SDE_LKFT}.

1) 
In Ref.~\cite{PhysRevD.66.105005}, within the assumption that in both three and four dimensions, the fermion propagator takes its free-particle form, Bashir and Raya used the LKFT to determine the propagator functions in any other covariant gauge. Their results can be reproduced by Eq.~\eqref{eq:LKFT_linearity_spectral_rep} when $\rho_1(s;0)=\delta(s-m^2)$ and $\rho_2(s;0)=m\delta(s-m^2)$ with distributions $\mathcal{K}_{j}$ operating on $z^\beta$ given by Eq.~\eqref{eq:kn_zbeta}.

2) QED in 4-dimensions with the Gauge Technique of Delbourgo, Salam and Strathdee [14-17]
\nocite{PhysRev.130.1287,PhysRev.135.B1398,PhysRev.135.B1428,Delbourgo:1977jc}
is a useful illustration of the form of the operator elements $\Omega_{ij}$ of Eq.~\eqref{eq:fermion_prop_SDE_abs}. These are readily deduced in the quenched approximation, which corresponds to the gray circles in Fig. \ref{fig:DSE_fermion_rho_photon_mod} being removed. In the language of the fermion-photon vertex $\Gamma^\mu(k,p)$, the Gauge Technique generates the longitudinal Ball-Chiu vertex \cite{PhysRevD.22.2542} with additional transverse pieces. When ultraviolet divergences are isolated by dimensional regularization, $\Omega_{ij}(s,s';\xi)$ consists of $\delta$-functions and $\theta$-functions; specifically, we have
\begin{align}
\Omega_{11}(s,s';\xi)& =-\dfrac{3\alpha}{4\pi}\left[\left(C_{div}+\dfrac{4}{3}+\ln\dfrac{\mu^2}{s}\right)\delta(s-s')-\dfrac{s'}{s^2}\theta(s-s') \right]\nonumber\\
& \quad -\dfrac{\alpha\xi}{4\pi}\dfrac{1}{s}\theta(s-s'),\nonumber\\[1mm]
\Omega_{12}(s,s';\xi)& =-\dfrac{m_B}{s}\delta(s-s'),\quad \Omega_{21}(s,s';\xi)=-m_B \delta(s-s'),\nonumber\\[1mm]
\Omega_{22}(s,s';\xi)&=-\dfrac{3\alpha}{4\pi}\left[\left(C_{div}+\dfrac{4}{3}+\ln\dfrac{\mu^2}{s}\right)\delta(s-s')-\dfrac{1}{s}\theta(s-s') \right]\nonumber\\
& \quad -\dfrac{\alpha\xi}{4\pi}\dfrac{s'}{s^2}\theta(s-s'),\label{eq:Omega_GT}
\end{align}
where $m_B$ is the bare mass and $C_{div}=1/\epsilon-\gamma_E+\ln 4\pi$. The $\delta$-function terms are the analogue of modifications to the propagator renormalization constants in perturbation theory, while the $\theta$-function terms correspond to corrections to the propagator from real particle production in the timelike region. 

The Gauge Technique in the quenched approximation is known to be inconsistent with LKFT \cite{Delbourgo:1980vc,Delbourgo:1979eu}. This can be seen just by inspecting the $\Omega_{21}$ component of Eq.~\eqref{eq:consistency_fermion_prop_SDE_LKFT}. Since $\mathcal{K}_2^{-\xi}\mathcal{K}_1^{\xi}\neq 1$, the requirement in this component is not met. 
Meanwhile, for small $\epsilon$, the operations given by Eq. \eqref{eq:kn_zbeta} can be written as
\begin{align}
\mathcal{K}_j=& \left(\dfrac{\mu^2z}{p^2} \right)^{-\nu}\exp\Bigg\{-\nu\left[\dfrac{1}{\epsilon}+\gamma_E+\ln 4\pi+\mathcal{O}(\epsilon^1) \right] \Bigg\}\times\nonumber\\
& z^{2-n}I^{\nu}z^{n-1-\nu}I^{\nu}z^{-\nu-1}.\label{eq:kn_small_epsilon}
\end{align}
one can verify that no component of Eq.~\eqref{eq:consistency_fermion_prop_SDE_LKFT} is satisfied by Eq.~\eqref{eq:Omega_GT}.


\section{Conclusions\label{ss:conclusion}}
In this article, we started with the structure of the fermion propagator using a spectral representation, which uniquely determines the propagator function in the complex momentum plane. This allows the LKFT for the fermion propagator spectral functions $\rho_j(s;\xi)$ to be solved exactly by keeping the number of spacetime dimensions explicit. Recognizing the vertex structure $S_F(k)\,\Gamma^\mu(k,p)\,S_F(p)$ is linear in $\rho_{j}(s;\xi)$, we then deduced an abstract version of the Schwinger-Dyson equation for the fermion propagator. Finally we derived the requirement for solutions of the fermion SDE in different covariant gauges to be consistent with LKFT in any dimensions. This can be used as a new criterion for truncating SDEs. This is clear if the ansatz is to hold in any covariant gauge. However, even if we restrict ourselves to solving the SDEs in one gauge, the ansatz should not change significantly with an infinitesimal change in gauge. Then the $\xi$-derivative of Eq.~\eqref{eq:consistency_fermion_prop_SDE_LKFT} must hold in that gauge.

Detailed discussion of solutions to the LKFT in a spectral representation can be found in Ref.~\cite{LKFT.fermion.epsilon}. 
In \cite{GC.propagators.SDEs} we 
make explicit those contributions to Eq.~\eqref{eq:consistency_fermion_prop_SDE_LKFT} that are exactly known without model truncations.
\section*{Acknowledgments}
This material is based upon work supported by the U.S. Department of Energy, Office of Science, Office of Nuclear Physics under contract DE-AC05-06OR23177 that funds Jefferson Lab research. The authors would like to thank Professor Keith Ellis and other members of the Institute for Particle Physics Phenomenology (IPPP) of Durham University for their kind hospitality during their visit when this article was finalized.


\bibliographystyle{elsarticle-num} 
\bibliography{GC_SDE_QED_arxiv_bib}





\end{document}